\documentclass[aps,pra,reprint,groupedaddress,amsmath,amssymb,longbibliography]{revtex4-2}

\usepackage{graphicx}
\usepackage{dcolumn}
\usepackage{bm}
\usepackage{xcolor}
\usepackage{braket}

\begin{document}

\title{Ideal Two-Color Field Ratio for Holographic Angular Streaking of Electrons}

\author{D. Trabert$^1$} 
\author{A. Geyer$^1$}
\author{N. Anders$^1$}
\author{M. Hofmann$^1$}
\author{M. S. Schöffler$^1$}
\author{L. Ph. H. Schmidt$^1$}
\author{T. Jahnke$^2$}
\author{M. Kunitski$^1$}
\author{R. Dörner$^1$}
\author{S. Eckart$^1$}
\email{eckart@atom.uni-frankfurt.de}

\affiliation{$^1$ Institut f\"ur Kernphysik, Goethe-Universit\"at, Max-von-Laue-Str. 1, 60438 Frankfurt am Main, Germany\\
$^2$Max-Planck-Institut f\"ur Kernphysik, Saupfercheckweg 4, 69117 Heidelberg, Germany 
}

\date{\today}
\begin{abstract}
We study strong field ionization of molecular hydrogen in highly intense co-rotating two-color (CoRTC) laser fields. The measured electron momentum distributions show alternating half-rings (AHR) that are characteristic for sub-cycle interference. We report on the role of the two-color field ratio for the visibility of this sub-cycle interference. The ratio of the peak electric field at 780 nm compared to the peak electric field at 390 nm $E_{780}/E_{390}$ is varied from 0.037 to 0.18. We find very good agreement with the results from our semiclassical simulation. We conclude that the AHR pattern is visible if two conditions are fulfilled. First, the amplitudes of the two pathways that lead to the sub-cycle interference have to be similar, which is the case for low two-color field ratios $E_{780}/E_{390}$. Second, the phase difference of the two pathways must be strong enough to allow for destructive interference, which is the case for high two-color field ratios $E_{780}/E_{390}$. For typical experimental conditions, we find that two-color field ratios $E_{780}/E_{390}$ in the range from 0.037 to 0.12 lead to good visibility of the AHR pattern. This guides future experiments to measure the Wigner time delay using holographic angular streaking of electrons (HASE).
\end{abstract}

\maketitle
\section{Introduction} 
When a highly intense laser pulse hits a single atom or molecule, an electron can be liberated by strong field tunnel ionization \citep{ADK86}. The outgoing electronic wave packet can be described by a complex valued wave function and carries information about the corresponding electronic bound state it emerged from \cite{Staudte2009,Eckart_HASEtheo,Trabert2021,Meckel2008} and the ionization dynamics \cite{Arissian2010,Eckart2018Banana,Trabert2021atomicH,kneller2022look}. The absolute square of the electronic  wave function in momentum space is routinely measured \cite{Ullrich_2003}. In contrast, a full characterization of the electronic wave packet requires to measure not only the electron wave packet's amplitude but also its phase. A very intriguing quantity that is closely related to the phase of the wave function is the Wigner time delay which is defined as the derivative of the phase of the wave function with respect to the electron's energy \cite{Wigner1954,rist2021measuring}. In the single- and two-photon regime, Wigner time delays were measured using RABBITT (reconstruction of attosecond beating by interference of two-photon transitions) \cite{Paul2001,Muller2002,vos2018orientation}. Recently, holographic angular streaking of electrons (HASE) has been introduced and it was demonstrated that it is the strong-field analogon of RABBITT. Accordingly, HASE allows for the measurement of Wigner time delays in the strong field regime \cite{Eckart_HASEtheo,Trabert2021,ArgonHASE}. This technique has been  used previously to measure angle-resolved Wigner time delays in molecular photoionization processes \cite{Trabert2021,ArgonHASE}. 
HASE experiments are carried out using a $\omega$-2$\omega$ bicircular co-rotating two-color (CoRTC) field that is composed of a strong $2\omega$ component and a weak $\omega$ component. While RABBITT uses two-photon transitions and is modeled via a two-path interference in the energy-domain, HASE is applied in the strong field regime and is modeled via a two-path interference in the time-domain. In previous works on HASE, the existence of the corresponding two-path sub-cycle interference in CoRTC fields has been demonstrated both theoretically and experimentally \cite{Han2018,Ge2019,Eckart_HASEtheo,Eckart2020_sideband}. The characteristic interference pattern observed in the photoelectron momentum distribution is known as the alternating half-ring (AHR) pattern and is a result of the  interference between two electronic wave packets that are released within two halfcycles of the $\omega$-comonent of the CoRTC field \cite{Eckart2020_sideband}. In the current work we show that two conditions are necessary to observe the AHR pattern. First, the two interfering wave packets must have similar amplitudes. Second, the phase difference for the sub-cycle interference must reach values up to $\pi$ (otherwise destructive interference is not possible). The detailed mechanism leading to the AHR pattern in the electron momentum distribution has been already successfully described by holographic angular streaking of electrons (HASE) \cite{Eckart2020_sideband,Eckart_HASEtheo}. The key-observable in HASE experiments is the rotation angle of the AHR pattern in the plane of polarization. Measuring this rotation angle allows for the retrieval of the electronic wave packet's phase gradient perpendicular to the tunnel exit in momentum space representation. The ability to measure this phase gradient is one of the key achievements of HASE and allows one to experimentally access Wigner time delays \cite{SimonPhD,Eckart_HASEtheo,Trabert2021,ArgonHASE}. 

The aim of this joint experimental and theoretical work is to obtain a better and more detailed understanding of the underlying two-path interference leading to the emergence of the AHR pattern in the electron momentum distribution. To this end, we study the strong field ionization of H$_{2}$ in CoRTC fields dominated by the $2\omega$ component at a central wavelength of 390\,nm and vary the intensity of the field at 780\,nm. The basic principles that lead to the sub-cycle interference are discussed in Sec. \ref{sec:CoRTC}. In Sec. \ref{sec:experiment}, the experimental techniques for light field synthesis and electron detection are described. We also describe how the intensity of the $\omega$ component at a central wavelength of 780\,nm is varied (which modifies the two-color field ratio). The setup that is used to measure three-dimensional momentum distributions of all charged reaction fragments in coincidence is presented and the measured electron momentum distributions are discussed. The section is concluded by a detailed analysis of the AHR pattern's visibility as a function of the two-color field ratio. In Sec. \ref{sec:simulation}, we employ a trajectory-based semi-classical model to reproduce the experimentally obtained electron momentum distributions and to gain insight into the sub-cycle phase-structure of the electronic wave function as well as the amplitude-ratio of the two wave packets that contribute to the sub-cycle interference. In Sec. \ref{sec:H2}, the rotation angle of the AHR as a function of the electron emission direction relative to the internuclear axis of H$_{2}$ is investigated for different two-color field ratios. Atomic units (a.u.) are used throughout this work unless stated otherwise.

\section{\label{sec:CoRTC}Strong field ionization in co-rotating two-color fields} 
The shape of a two-path interference pattern depends on the amplitudes and the relative phase of the two interfering wave packets and can be expressed as:
\begin{equation}
\label{eq:doubleslit}
I\left(\Delta\Phi\right) = \left|A_{1}\mathrm{e}^{\mathrm{i}\Phi_{1}}+A_{2}\mathrm{e}^{\mathrm{i}\left(\Phi_{1}+\Delta\Phi\right)}\right|^{2}
\end{equation}
Here, the intensity $I$ depends on the relative phase $\Delta\Phi$ of the two waves and their amplitudes ($A_{1}$ and $A_{2}$). The absolute phase $\Phi_{1}$ does not affect the interference pattern. The modulation-strength of the interference pattern is governed by the ratio of the amplitudes $A_{1}$ and $A_{2}$.

\begin{figure}[h]
\includegraphics[width=\columnwidth]{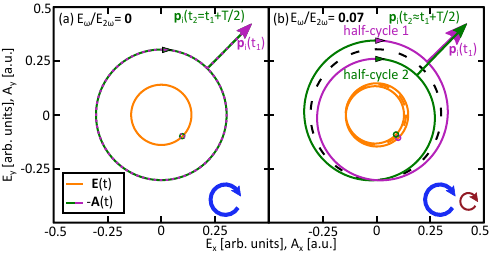}
\caption{\label{fig1} Electric field and negative vector potential of a single color and  a CoRTC field. (a) Lissajous curve for the electric field $\mathbf{E}(t)$ and the corresponding negative vector potential $-\mathbf{A}(t)$ for a circularly polarized $2\omega$ light field with $E_{2\omega}=0.035$\,a.u. Each point in final momentum space can be reached by a well-defined initial momentum $\mathbf{p}_{\mathrm{i}}$. $t_{1}$ and $t_{2}$ fulfill the relation $t_{2}= t_{1}+T/2$. (b) Time-dependent electric field $\mathbf{E}(t)$ and negative vector potential $-\mathbf{A}(t)$ for a CoRTC field with a two-color field ratio $\frac{E_{\omega}}{E_{2\omega}}=0.07$. Each point in final momentum space can be reached by two combinations of birth time $t_{1,2}$ and initial momentum $\mathbf{p}_{\mathrm{i}}\left(t_{1,2}\right)$. $t_{1}$ and $t_{2}$ approximately fulfill the relation $t_{2}\approx t_{1}+T/2$. The negative vector potential that belongs to half-cycle 1 [2] is represented by the purple [green] line. The electric field vectors that belong to the release times within the first [second] half-cycle are indicated in purple [green].}
\end{figure}

Fig. \ref{fig1} illustrates the temporal evolution of the electric field $\mathbf{E}(t)$ and the negative vector potential $-\mathbf{A}(t)$ in the plane of polarization ($xy$-plane) for a single-color circularly polarized $2\omega$ field (Fig. \ref{fig1}(a)) and a CoRTC field (Fig. \ref{fig1}(b)). The green and the purple arrows in Fig. \ref{fig1}(b) visualize that each point in final momentum space can be reached by two combinations of negative vector potential and initial momentum at the tunnel exit. To understand why this is the case, one can describe the ionization as a two-step process: First, the electron tunnels through the classically forbidden region which results from the binding potential deformed by the instantaneous laser electric field. Second, the electron is subsequently accelerated in the combined electric field of the laser and the parent ion. Neglecting Coulomb interaction with the parent ion after tunneling, an electron which appears at the tunnel exit at a time $t_{0}$ reaches a final momentum $\mathbf{p}_{\mathrm{elec}}=-\mathbf{A}(t_{0})+\mathbf{p}_{\mathrm{i}}$. Here, $-\mathbf{A}(t_{0})$ is the negative vector potential at the time $t_{0}$ and $\mathbf{p}_{\mathrm{i}}$ is the initial momentum at the tunnel exit. For circularly polarized single-color fields it is often assumed that the component of  $\mathbf{p}_{\mathrm{i}}$ in tunnel-direction is negligible \cite{Ni2016}. With this assumption, there remain only two possible combinations of $-\mathbf{A}(t_{0})$ and $\mathbf{p}_{\mathrm{i}}$ to reach a certain point in final momentum space  \cite{Eckart_HASEtheo}.

Fig. \ref{fig1}(a) shows the limiting case of a CoRTC field with a vanishing $\omega$-component (i.e. a single color $2\omega$ field). Within this limiting case, the two half-cycles of the ``CoRTC field'' are exactly identical. Thus, there are two release times $t_{1}$ and $t_{2}=t_{1}+T/2$ per cycle $T$ that lead to the same final electron momentum $\mathbf{p}_{\mathrm{elec}}$. The initial momenta at the tunnel exit $\mathbf{p}_{\mathrm{i}}\left(t_{1}\right)=\mathbf{p}_{\mathrm{i}}\left(t_{2}\right)$ and the negative vector potentials $-\mathbf{A}(t_{1})=-\mathbf{A}(t_{2})$ are identical (illustrated by the green and purple arrows in Fig. \ref{fig1}(a)). For this case, the phase difference between the two paths leading to $\mathbf{p}_{\mathrm{elec}}$ is $\frac{1}{\hbar}I_{\mathrm{p}}T/2$, where $I_{\mathrm{p}}$ denotes the ionization potential \cite{Shvetsov-Shilovski2016}. Further, the tunneling probability for any of the two half-cycles is identical, corresponding to the special case $A_{1}=A_{2}$ in Eq. \ref{eq:doubleslit} and thus allowing for perfect constructive and destructive interference in the final electron momentum distribution. In this case, the phase difference gives rise to interference maxima that are well-known as above-threshold-ionization (ATI) peaks in the photoelectron energy distribution and do not produce any AHR pattern because the phase difference is independent of the electron release time and thus detection angle.

Fig. \ref{fig1}(b) shows the combined electric field for a two-color field ratio $\frac{E_{\omega}}{E_{2\omega}}$ of 0.07, which is typical for previous experimental works \cite{Eckart2020_sideband,Trabert2021,ArgonHASE}. For the two half-cycles with a duration of $T/2$, the electric field and the corresponding negative vector potential are not identical. In such a laser field, with a non-zero $\omega$ component, there are three major differences to the situation in a circularly polarized single-color $2\omega$ field: First, the geometry is more complicated, since $-\mathbf{A}(t_{0})$ is not perpendicular to $\mathbf{E}(t_{0})$ for all release times $t_{0}$ anymore. The time between release time $t_{2}$ in the second half-cycle and the release time $t_{1}$ in the first half-cycle is now only approximately $t_{1}+T/2$. The initial momentum which is needed to end up at a certain final electron momentum $\mathbf{p}_{\mathrm{elec}}$ is different for the two half-cycles (which is due to the differences of the negative vector potential). Second, the deviations in the relative release time from the exact $T/2$ periodicity and the difference in initial momentum lead to a more complicated evolution of the semi-classical phase. Thus, $\Delta\Phi$ can deviate from $\frac{I_{\mathrm{p}}T}{2\hbar}$. Third, the ratio of amplitudes of the two interfering wave packets changes because $\mathbf{E}(t)$ varies on  sub-cycle time scale and the tunneling probability depends on the magnitude of the laser electric field. This difference in amplitudes reduces the modulation-strength of the interference pattern according to Eq. \ref{eq:doubleslit} and is illustrated by the different thicknesses of the green and purple arrows representing $p_i$ in Fig. \ref{fig1}(b). 

\begin{figure}[h]
\includegraphics[width=\columnwidth]{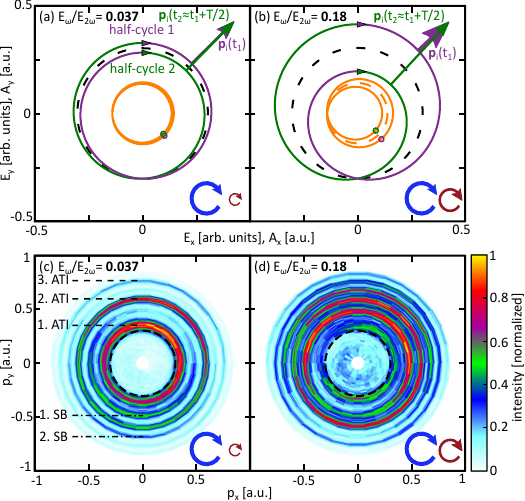}
\caption{\label{fig2} Electric field and negative vector potential of CoRTC fields with varying two-color field ratios and corresponding measured electron momentum distributions. (a) Time-dependent electric field $\mathbf{E}(t)$ and negative vector potential $-\mathbf{A}(t)$ of a CoRTC field with $E_{2\omega}=0.035$\,a.u. and a two-color field ratio of $\frac{E_{\omega}}{E_{2\omega}}=0.037$. (b) shows the same as (a) but for a two-color field ratio of $\frac{E_{\omega}}{E_{2\omega}}=0.18$. The negative vector potential of half-cycle 1 [2] is represented by the purple [green] line. The electric fields corresponding to the release times within the first [second] half-cycle are indicated by purple [green] circles. (c) Measured electron momentum distribution in the plane of polarization ($xy$-plane) for $\frac{E_{\omega}}{E_{2\omega}}=0.037$. A weakly modulated AHR pattern is visible. The ATI and SB peaks are labeled. (d) shows the same as (c) but for $\frac{E_{\omega}}{E_{2\omega}}=0.18$. Here, the AHR pattern is less pronounced. The dashed circles in all panels show the negative vector potential for the circularly polarized single-color field shown in Fig. \ref{fig1}(a).}
\end{figure}

In order to unravel how the field ratio $\frac{E_{\omega}}{E_{2\omega}}$ effects the AHR pattern, different field ratios are compared. Fig. \ref{fig2}(a) [(b)] shows the time-dependent electric field and negative vector potential of the smallest [largest] two-color field ratio of 0.037 [0.18] investigated in this work. While the splitting between the two half-cycles is very small for the field depicted in Fig. \ref{fig2}(a), the field shown in Fig. \ref{fig2}(b) has a very large splitting and the ionization process can be expected to be dominated by the half-cycle with the larger field amplitude. The measured electron momentum distributions in the plane of polarization ($xy$-plane) are shown in Fig. \ref{fig2}(c) and (d) (see Sec. \ref{sec:experiment} for experimental details). In Fig. \ref{fig2}(c), the well-known AHR pattern can be clearly seen. In contrast, Fig. \ref{fig2}(d) shows that the AHR pattern can hardly been seen for the field ratio of $\frac{E_{\omega}}{E_{2\omega}}=0.18$. In the remainder of this work, we will discuss why the AHR pattern is not seen for very low ratios $\frac{E_{\omega}}{E_{2\omega}}$, appears for ratios on the order of 0.037 and vanishes again for high field ratios.

\section{\label{sec:experiment}Experimental details and analysis of measured electron momentum distributions} 
\begin{figure*}[ht]
\includegraphics[width=1\textwidth]{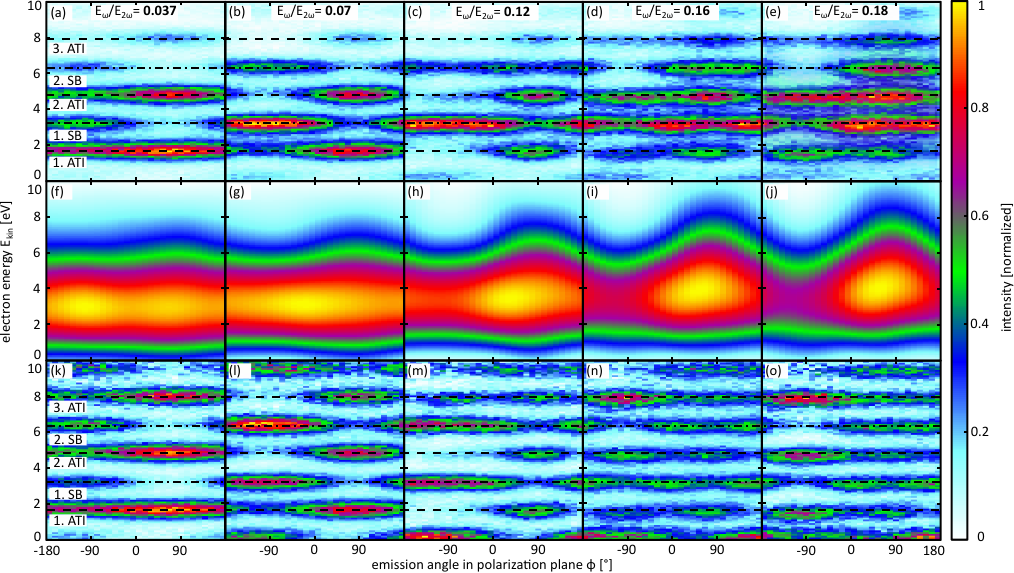}
\caption{\label{fig3} Measured electron energy distributions for varying CoRTC field ratios $\frac{E_{\omega}}{E_{2\omega}}$. (a)-(e) measured electron momentum distributions as a function of the angle $\phi=\mathrm{atan2}\left(p_{y},p_{x}\right)$ and the energy $E_{\mathrm{kin}}=\frac{p_{x}^{2}+p_{y}^{2}+p_{z}^{2}}{2}$ for two-color field ratios $\frac{E_{\omega}}{E_{2\omega}}$ from 0.037 to 0.18. The three ATI [two SB] peaks visible are indicated by dashed [dashed-dotted] lines. (f)-(j) show envelopes (low-pass filtered distribution to remove the modulation due to the photon energy, see text) extracted from the measured distributions shown in (a)-(e). For larger field ratios, a stronger modulation of the angular distribution is observed. (k)-(o) interference patterns obtained by dividing the measured distributions shown in (a)-(e) by the corresponding envelopes shown in (f)-(j) for each bin separately. The emergence and the breakdown of the AHR pattern can be observed. (l) shows the most symmetric AHR pattern.}
\end{figure*}

To generate the CoRTC fields used in our experiment, laser pulses at a fundamental frequency $\omega$ (wavelength of 780\,nm) are frequency doubled using a 200\,\textmu m $\beta$-barium borate crystal, resulting in laser pulses at a frequency $2\omega$. The setup is the same as in \cite{Eckart2016,Trabert2021,Eckart2020_sideband}. A dichroic beamsplitter separates the two frequency components such that the intensity and the polarization of the $\omega$ and the $2\omega$ pulses can be set independently. A piezo delay stage allows us to ensure temporal overlap and to control the relative phase of the two single-color laser pulses that form the CoRTC field. The resulting two-color laser pulses are focused by a spherical mirror (focal length $f=75$\,mm) onto a cold supersonic jet of hydrogen molecules resulting in an intensity (peak electric field) of $8.4 \times 10^{13}$\,W/cm$^{2}$ ($E_{2\omega}=0.035$\,a.u.) for the $2\omega$ component of the CoRTC field. A gradient neutral density filter is used to set 12 different intensities (peak electric fields) for the $\omega$ component ranging from $1.4 \times 10^{11}$\,W/cm$^{2}$ ($E_{\omega}=0.0014$\,a.u.) to $2.8 \times 10^{12}$\,W/cm$^{2}$ ($E_{\omega}=0.0064$\,a.u.). This leads to two-color field ratios between $\frac{E_{\omega}}{E_{2\omega}}=0.037$ and $\frac{E_{\omega}}{E_{2\omega}}=0.18$. The intensity calibration is done as in \cite{Trabert2021}. After ionization, electrons and ionic fragments are guided by static electric and static magnetic fields towards the position- and time-sensitive detectors of a cold target recoil ion spectroscopy (COLTRIMS) reaction microscope \cite{Dorner2000}. Each detector is comprised of a double stack of micro channel plates and hexagonal delayline-anodes \cite{Jagutzki_2002}. The ion [electron] arm of the spectrometer has a length of 66\,mm [390\,mm], an electric field of 10.9\,V/cm and a magnetic field of 8.1 Gauss. The three-dimensional momentum vectors of all charged particles are measured in coincidence. For the current work, single ionization followed by dissociation of the hydrogen molecule is studied. Each event in the electron momentum distributions has been measured in coincidence with a single H$^{+}$ ion released via the reaction $\mathrm{H}_{2}\rightarrow\mathrm{H}+\mathrm{H}^{+}+e^{-}$.

To further analyze electron momentum distributions as in Fig. \ref{fig2}(c) and \ref{fig2}(d), we calculate the 
electron's kinetic energy $E_{\mathrm{kin}}=\frac{p_{x}^2+p_{y}^2+p_{z}^2}{2}$ and its emission angle in the plane of polarization $\phi=\mathrm{atan2}\left(p_{y},p_{x}\right)$. Fig. \ref{fig3}(a)-\ref{fig3}(e) show the measured electron energy distributions as a function of these two quantities for some of the measured two-color field ratios $\frac{E_{\omega}}{E_{2\omega}}$. The three ATI peaks and the two SB peaks that are relevant in the further discussion are indicated by the dashed and dashed-dotted lines. In analogy to Ref. \cite{Trabert2021} we disentangle the observed pattern into a product of an envelope function (Fig. \ref{fig3}(f)-\ref{fig3}(j)) and the underlying AHR pattern (Fig. \ref{fig3}(k)-\ref{fig3}(o)). The envelope function is obtained by suppressing Fourier components in the energy distribution corresponding to the modulation that is due to the photon energy and higher frequencies while ensuring that the integral for each bin along $\phi$ does not change. This produces the experimentally obtained envelopes that are depicted in Fig. \ref{fig3}(f)-\ref{fig3}(j). The modulation-strength of the angular distribution increases for higher intensities of the $\omega$ component as expected due to the non-linearity of the tunneling probability as a function of the mangnitude of the laser's electric field. For the field ratios 0.12, 0.16 and 0.18 (Fig. \ref{fig3}(h)-\ref{fig3}(j)), the most probable kinetic energy increases due to the increasing magnitude of the negative vector potential of the dominant half-cycle. The experimental data have been rotated in the plane of polarization such, that the most probable electron emission angle (integrated over the electron's energy) qualitatively agrees with the one obtained from the semi-classical simulation (see Sec. \ref{sec:simulation}). For the simulation, the maximum splitting in the negative vector potential is at $\phi=90^{\circ}$ as illustrated in Fig. \ref{fig1}(b) and Fig. \ref{fig2}(a) and \ref{fig2}(b).

For a two-color field ratio of 0.037 (Fig. \ref{fig3}(k)), the SB peaks are only weakly populated but show a strong angular modulation. For a two-color field ratio of 0.07 (Fig. \ref{fig3}(l)), the ATI and SB peaks are equally populated and show a similar angular modulation-strength. This is the two-color field ratio used in previous works that applied the HASE technique to experimentally investigate the Wigner time delay \cite{Trabert2021,ArgonHASE}. Higher field ratios as in Fig. \ref{fig3}(l)-\ref{fig3}(o) increase the population of the SB peaks compared to the ATI peaks.   

\begin{figure}[h]
\includegraphics[width=1\columnwidth]{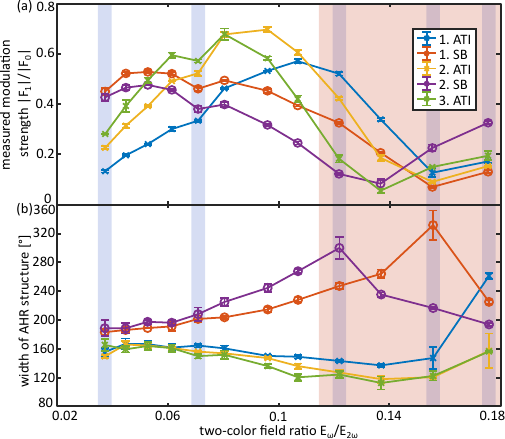}
\caption{\label{fig4} Measured modulation strength (AHR visibility) and width of the AHR pattern. (a) Shows the visibility of the AHR pattern as a function of the two-color field ratio $\frac{E_{\omega}}{E_{2\omega}}$. Here the visibility is expressed as the modulation-strength which in turn is given by the ratio between the one-fold symmetric component $F_{1}$ and the DC component $F_{0}$ of the discrete Fourier transform of the angular distribution for each energy peak. (b) Width (FWHM, see text) of the AHR pattern for each energy peak as a function of the two-color field ratio $\frac{E_{\omega}}{E_{2\omega}}$. Data points corresponding to ATI [SB] peaks are indicated by crosses [circles]. The data points within the blue shaded areas indicate the field ratios shown in more detail in Fig. \ref{fig3}. The red area highlights field ratios for which the AHR pattern shows more than a single peak (but also double peaks, e.g. for the 1st SB of Fig. \ref{fig3}(n)). Error bars show the statistical error only.}
\end{figure}

In order to conduct a more quantitative discussion we calculate the discrete Fourier transform of the angular distributions as in Fig. \ref{fig3}(k)-(o) for each energy peak. Fig. \ref{fig4}(a) shows the ratio of the one-fold symmetric component $F_{1}$ and the integral DC component $F_{0}$. For a circularly polarized single-color pulse, as illustrated in Fig. \ref{fig1}(a), this quantity would be zero. Fig. \ref{fig4}(a) shows that for the smallest two-color field ratio of 0.037, the one-fold symmetric modulation is much stronger for the SB than for the ATI peaks. The reason for this is obvious when looking at Fig. \ref{fig3}(k). While the modulation of the SB peaks is ideal for small two-color field ratios, the angular distributions corresponding to the ATI peaks remain rather isotropic. For increasing two-color field ratios, the Fourier component $F_{1}$ increases for the ATI peaks and eventually reaches its maximum at slightly different field ratios for the individual energy peaks. For the two SB peaks, $F_{1}/F_{0}$ remains rather constant up to a two-color field ratio of around 0.1 and then decreases rapidly. The red area in Fig. \ref{fig4} highlights two-color field ratios for which the angular distributions for the energy peaks are not dominated by single peaks (e.g. as for the 1st SB of Fig. \ref{fig3}(n)). 

The difference in the intensity comparing the ATI versus the SB peaks in Fig. \ref{fig4}(a) is in qualitative agreement with the expectation from the photon picture. In the photon picture (i.e. in the energy domain that is used to interpret RABBITT experiments), the AHR pattern on the SB is caused by an interference between two channels. One channel for which N photons at $2\omega$ are absorbed and one photon at $\omega$ is absorbed and the other channel for which N+1 photons at $2\omega$ are absorbed and one photon at $\omega$ is emitted (stimulated emission). As soon as the transition amplitude for absorption and emission of one $\omega$ photon becomes appreciable, the sidebands arise in an energy region that was empty before.  The increased intensity on the sideband manifests as a depletion for the main energy peak that belongs to a certain ATI peak. For these main energy peaks the modulation is often small compared to the signal of the entire peak (that also exists without the $\omega$ field). Although the photon picture can be used to qualitatively understand certain features of the AHR pattern, it is complicated to model all phase shifts in the energy domain and relies on the assumption that the $\omega$ field can be treated perturbatively \cite{Zipp2014}. The semi-classical approach, that is used for the remainder of this paper, does not rely on this assumption and reproduces the AHR pattern as a two-path interference in the time-domain \cite{Eckart_HASEtheo}. Thus, we will use this semi-classical time-domain modeling (section IV) which fully consistently models our findings without the need to invoke the photon picture (energy domain perspective) as done in RABBITT.



Fig. \ref{fig4}(b) shows the angular width of the AHR pattern as a function of the two-color field ratio for each energy peak. To obtain this quantity, the one-dimensional angular distributions were low-pass filtered for each energy peak. There are usually two angles for which these smoothed curves yield half of the maximum modulation-strength as a function of the emission angle. The distance between those angles is shown in the figure as the width of the AHR pattern and corresponds to the full width at half maximum (FWHM). For low two-color field ratios, it is close to 180$^{\circ}$. For the SB peaks, the width increases with the field ratio, while for the ATI peaks, the width decreases. This effect can be clearly seen comparing Fig. \ref{fig3}(k) and Fig. \ref{fig3}(m). Our results are in line with previous theoretical works investigating linearly polarized two-color fields \cite{Feng2019}. Note that for field ratios that are within the red area in Fig. \ref{fig4}, the peak width cannot be unambiguously determined.
 
\section{\label{sec:simulation}Investigation of two-path interference using our semi-classical model} 
\begin{figure*}[ht!]
\includegraphics[width=1\textwidth]{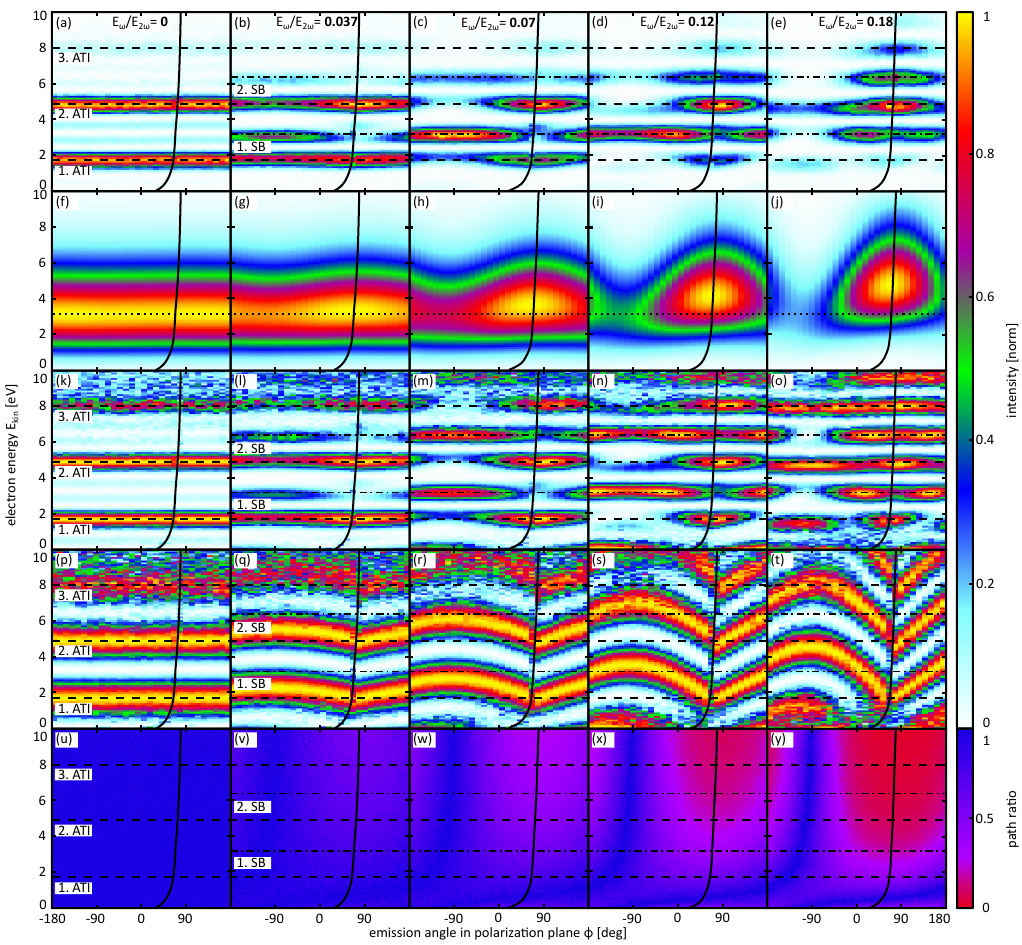}
\caption{\label{fig5} Analysis of the two-path, sub-cycle interference using the semi-classical model. (a)-(e) show the results from the full semi-classical simulations as a function of $E_{\mathrm{kin}}$ and $\phi$ for two-color field ratios from 0 to 0.18 as indicated in the panels. In analogy to Eq. \ref{eq:doubleslit} this can be expressed via $\left|\sum_{n=1}^{4}A_{n}\mathrm{exp}\left(\mathrm{i}\Phi_{n}\right)\right|^{2}$. (f)-(j) shows the same as (a)-(e) but here the four simulated electron trajectories, that are launched during all four half-cycles, are added up incoherently to suppress interference, which can be written as $\sum_{n=1}^{4}|A_{n}|^2 $. The dotted horizontal line indicates the most probable energy for the single-color $2\omega$ field with an peak electric field of 0.035\,a.u. (k)-(o) Normalized coherent sum which can be expressed as $\frac{\left|\sum_{n=1}^{4}A_{n}\mathrm{exp}\left(\mathrm{i}\Phi_{n}\right)\right|^{2}}{\sum_{n=1}^{4}|A_{n}|^2 }$ in full analogy to the row above. The ATI [SB] peaks are indicated by the dashed [dashed-dotted] lines. (p)-(t) Sub-cycle interference pattern $\left|\mathrm{exp}\left(\Phi_{1}\right)+\mathrm{exp}\left(\Phi_{2}\right)\right|^{2}$ between the first two half-cycles. (u)-(y) Path ratio $1-\left|\frac{A_{1}-A_{2}}{A_{1}+A_{2}}\right|$ for the sub-cycle two-path interference. The curved black lines indicate the transition between half-cycle 1 [3] and 2 [4].}
\end{figure*}

To better understand the origin of the two-path interference that gives rise to the AHR pattern for CoRTC fields, we use a semi-classical trajectory-based model \cite{Brennecke2020}. It is similar to the semi-classical two-step (SCTS) model \cite{Shvetsov-Shilovski2016} but uses initial conditions at the tunnel exit from strong-field approximation (SFA) \cite{Popruzhenko} as in \cite{Trabert2021atomicH,Geyer2022} but also includes the semi-classical phase associated to each classical trajectory which enables the modeling of interference in final momentum space. 

For our semi-classical simulations, we include Coulomb interaction after tunneling between the electron and the parent ion. It is important to note, that SFA predicts a non-zero initial momentum component parallel to the tunnel exit direction, when the magnitude of the laser's electric field changes as a function of time (as is the case for CoRTC fields with a non-vanishing $\omega$ component). For our simulations, we use a two-cycle flat-top CoRTC pulse with a peak electric field of 0.037\,a.u. for the $2\omega$ component and field ratios of 0, 0.037, 0.07, 0.12, and 0.18. The peak electric field of the $2\omega$ component is chosen such that the envelope in Fig. \ref{fig5}(g) has the same most probable energy as in Fig. \ref{fig3}(f). The ionization potential $I_{\mathrm{p}}$ of 16.17\,eV  was used, which leads to a good agreement of the energy peaks in Fig. \ref{fig5}(l) with the peaks in Fig. \ref{fig3}(k) and is consistent with the expectation for the non-Franck-Condon transition in strong-laser fields \cite{Dietrich1996,Posthumus2004}.

For the further analysis, the amplitudes $A_{1,2,3,4}$ and the phases $\Phi_{1,2,3,4}$ of the semi-classical electronic wave function in final momentum space are discussed for values $t_{1,2,3,4}$ of $t_{0}$ in the four half-cycles of the flat-top part of the pulse that is used for the simulation. This allows for the calculation of incoherent intensity distributions as well as for the modeling of sub-cylce and inter-cycle interference patterns in full analogy to the discussion in Ref. \cite{Eckart_HASEtheo}.

Fig. \ref{fig5}(a)-\ref{fig5}(e) shows the results of our semi-classical simulations for several field ratios in full analogy to the experimental results (see Fig. \ref{fig3}(a)-(e)). Based on Eq. 1 this can be expressed via $\left|\sum_{n=1}^{4}A_{n}\mathrm{exp}\left(\mathrm{i}\Phi_{n}\right)\right|^{2}$. We find good agreement of the semi-classical simulations with the experimental results. 

Fig. \ref{fig5}(f)-\ref{fig5}(j) shows the incoherent sum $\sum_{n=1}^{4}\left|A_{n}\right|^2$ of electron trajectories launched during all four half-cycles. The result can be compared with the envelopes extracted from the experimental data shown in Fig. \ref{fig3}(f)-\ref{fig3}(j). The angle-dependent energy distributions are in very good agreement with the experimental observations. The simulated modulation-strength of the angular distribution is slightly more pronounced in our simulation compared to the experiment. The narrower energy distribution for a given value of $\phi$ can be explained by the focal averaging which is inherently present in the experiment but is not included in our simulation. The curved black line is the same in all panels of Fig. \ref{fig5} and guides the eye indicating the transition from the first [third] to the second [forth] half-cycle. The line's curvature is a result of an energy-dependent attoclock offset angle which is due to Coulomb interaction between the electron and the parent ion after tunneling \cite{Eckle2008,Bray2018,Trabert2021atomicH}.

Fig. \ref{fig5}(k)-\ref{fig5}(o) show the normalized coherent sum $\frac{\left|\sum_{n=1}^{4}A_{n}\mathrm{exp}\left(\mathrm{i}\Phi_{n}\right)\right|^{2}}{\sum_{n=1}^{4}\left|A_{n}\right|^2 }$ over all four half-cycles in the flat-top part of the pulse. This is equivalent to the normalized measured interference patterns shown in Fig. \ref{fig3}(k)-\ref{fig3}(o). As before, the ATI [SB] peaks are indicated by dashed [dashed-dotted] lines. The simulated interference patterns are in very good qualitative agreement with their measured counterparts. 

In the next and final step of this section, the amplitudes and the phase structure of the sub-cycle wave packets are discussed. To this end, we consider one full cycle of the CoRTC comprised of the first half-cycle and the second half-cycle only.  Fig. \ref{fig5}(p)-\ref{fig5}(t) show the sub-cycle interference $\left|\mathrm{exp}\left(\Phi_{1}\right)+\mathrm{exp}\left(\Phi_{2}\right)\right|^{2}$ considering only the phase difference and neglecting amplitudes. For the case of a circularly polarized single-color laser field the CoRTC sub-cycle interference is just identical to the $2\omega$ inter-cycle interference, exhibiting ATI peaks spaced by the energy $2\hbar\omega$ independent of the angle $\phi$ (see Fig. \ref{fig5}(p)). For a non-zero $\omega$ component (see Fig. \ref{fig5}(q)-\ref{fig5}(t)), the interference pattern has the shape of arcs that extend further along the energy direction for increasing two-color field ratios. These arcs are the underlying reason for the observation of the AHR pattern \cite{Eckart_HASEtheo,Eckart2020_sideband}. Additionally, inter-cycle interference leads to discrete final electron energies (indicated by the dashed and dashed-dotted horizontal lines in Fig. \ref{fig5}(q)-(t)) and thus cuts out arc segments leading to the AHR pattern. For the smallest non-zero two-color field ratio of 0.037, the arcs are only weakly bent (see Fig. \ref{fig5}(l)) and hence lead to intense ATI peaks compared to relatively weak SB peaks. 

For larger two-color field ratios, the arcs are so strongly bend, that a two-fold symmetric angular distribution can emerge for certain energy peaks (see e.g. last column of Fig. \ref{fig5}). Although the sub-cycle interference for a two-color field ratio of 0.18 predicts destructive interference for the SB peaks around $\phi=90^{\circ}$ (see Fig. \ref{fig5}(t)), the full interference pattern shown in Fig. \ref{fig5}(e) has no clear minimum because the envelope dominates (see Fig. \ref{fig5}(j)). This gives a hint to the second important aspect in the formation of the AHR pattern: unequal amplitudes $A_{1}$ and $A_{2}$ limit the achievable modulation-strength of the sub-cycle interference. 

In order to quantify the ratio of the amplitudes of the two wave packets that lead to the sub-cycle interference, we define the path ratio as $1-\left|\frac{A_{1}-A_{2}}{A_{1}+A_{2}}\right|$. Accordingly, the  path ratio has the value of 1 if $A_{1}=A_{2}$ and the value of 0 if either $\frac{A_{1}}{A_{2}}$ or $\frac{A_{2}}{A_{1}}$ goes to zero. This quantity is shown in Fig. \ref{fig5}(u)-\ref{fig5}(y). For the single-color $2\omega$ field, $A_{1}$ and $A_{2}$ are independent of $E_{\mathrm{kin}}$ and $\phi$ (see Fig. \ref{fig5}(u)). For all other cases around $\phi=90^{\circ}$, where the splitting in the negative vector potential and the corresponding electric field between the two half-cycles is at its maximum, the path ratio reaches its minimum. For a two-color field ratio of 0.18, the path ratio is close to zero for a wide range of combinations of $E_{\mathrm{kin}}$ and $\phi$ (see Fig. \ref{fig5}(y)). The only exception is the region around $\phi=-90^{\circ}$, where the shape of the negative vector potential and the corresponding electric field is almost independent of the two-color field ratio and is similar to a single-color $2\omega$ field. This is supported by the energy spacing of $2\hbar\omega$ that can be observed around $\phi=-90^{\circ}$ in Fig. \ref{fig5}(o). At the same time, the electric field leading to a relevant ionization probability has an effective periodicity which is $\omega$, leading to an energy spacing of only $\hbar\omega$ (see $\phi=90^{\circ}$ in Fig. \ref{fig5}(e)). Thus, our semi-classical simulation provides a clear and quantitative understanding why the AHR pattern vanishes for low and high two-color field ratios. 

\section{\label{sec:H2}Rotation of the AHR pattern as a function of the electron emission direction relative to the internuclear axis of H$_{2}$} 
\begin{figure}[h]
\includegraphics[width=1\columnwidth]{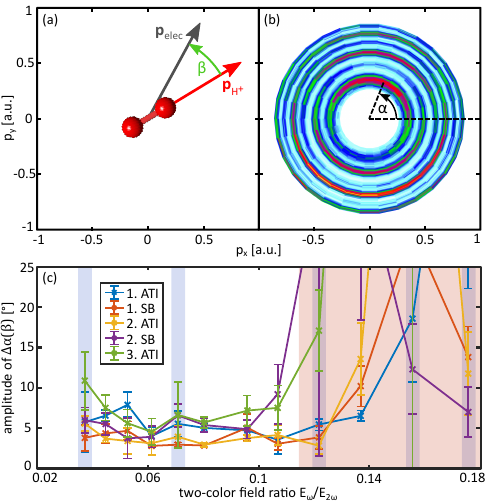}
\caption{\label{fig6} The rotation of the AHR pattern as a function of the two-color field ratio for the reaction $\mathrm{H}_{2}\rightarrow\mathrm{H}+\mathrm{H}^{+}+e^{-}$. (a) shows the definition of $\beta$ as the angle between the final electron momentum $\mathbf{p}_{\mathrm{elec}}$ and the momentum vector $\mathbf{p}_{\mathrm{H^{+}}}$ of the proton in the light's plane of polarization. (b) Definition of the angle $\alpha$ indicating the most probable electron emission direction for each ATI [SB] peak in the plane of polarization. (c) Amplitude of the sinusoidal dependence of $\Delta\alpha$ on $\beta$ as a function of the two-color field ratio. The blue areas indicate the field ratios shown in more detail in Fig. \ref{fig3} and are the same as in Fig. \ref{fig4}. The red area is the same as in Fig. \ref{fig4}. Error bars in (b) show the statistical error only.}
\end{figure}

In this section, it is analzed how a central observable in HASE experiments, namely the rotation angle $\alpha$ (see Fig. \ref{fig6}(a)), depends on the two-color field ratio. This rotation angle $\alpha$ allows one to measure the energy- and angle resolved Wigner time delays upon the strong field ionization of molecules \cite{Trabert2021}. Using the same approach and data analysis technique as in Ref. \cite{Trabert2021}, we divide the full data set into subsets for different relative angles $\beta$. Here, $\beta$ is the angle between the electron's momentum vector $\mathbf{p}_{\mathrm{elec}}$ and the proton's momentum vector $\mathbf{p}_{\mathrm{H}^{+}}$ for the reaction $\mathrm{H}_{2}\rightarrow\mathrm{H}+\mathrm{H}^{+}+e^{-}$ (see Fig. \ref{fig6}(a)). 

For a given value of $\beta$, the electron momentum distribution in the plane of polarization is obtained. After demerging the envelope and the interference pattern as for the third row of Fig. \ref{fig3} the rotation angle $\alpha$ for each ATI and SB peak can be determined (see Fig. \ref{fig6}(b)). This is done for all values of $\beta$ to obtain $\alpha\left(\beta\right)$. Subtracting the mean value $\alpha_{\mathrm{mean}}$ from $\alpha\left(\beta\right)$ for each ATI and SB peak one obatains the quantity $\Delta\alpha\left(\beta\right)$, which can be directly associated with phase gradients in momentum space perpendicular to the tunnel exit \cite{Eckart_HASEtheo,Trabert2021}. An approximately sinusoidal curve shape of $\Delta\alpha\left(\beta\right)$ is typically observed \cite{Trabert2021}. In a final step of this work, we quantify the influence of the two-color field ratio on the observable $\Delta\alpha\left(\beta\right)$. To this end, we determine the amplitude of the sinusoidal dependence as a function of the two-color field ratio. The result is shown in Fig. \ref{fig6}(c). Within the statistical uncertainty, the value remains constant for all field ratios outside of the red area (i.e. two-color field ratios at which the AHR is clearly visible). This leads to the conclusion, that the HASE technique works even if the field ratio is varied by a factor of 3 with respect to the two-color field ratio at which the AHR is most pronounced (corresponds to a change in intensity of the $\omega$ component by an order of magnitude).

\section{Conclusion}
In this joint experimental and theoretic study we have investigated the alternating half-ring (AHR) pattern observed for CoRTC fields dominated by the $2\omega$ component. We find that the AHR pattern is visible for two-color field ratios $E_{780}/E_{390}$ in the range from 0.037 to 0.12 for typical experimental conditions (ionization of molecular hydrogen and an intensity on the order of $10^{14}$W/cm$^2$). For higher field ratios the AHR pattern, which is a sub-cycle interference pattern, cannot be observed because only one of the two wave packets, that contributes to the sub-cycle interference, has a relevant amplitude. For very low field ratios, the phase differences for the two path interference is independent of the electron's release time and reproduces the well-known ATI peaks for a single color field at $2\omega$. Our findings show why a field ratio $E_{780}/E_{390}$ of 0.06 to 0.07 was chosen in previous experiments \cite{Trabert2021}. Finally, our results lead to a comprehensive and quantitative understanding of the AHR pattern and are of practical significance for HASE experiments, as they identify a region of two-color field ratios that allow for the application of the HASE technique and the retrieval of Wigner time delays.  

\section{Acknowledgments}
\normalsize
This work was supported by the DFG (German Research Foundation) through Priority Programme SPP 1840 QUTIF. M.H. acknowledges funding from the DFG – Project No. 328961117 – SFB 1319 ELCH (Extreme light for sensing and driving molecular chirality). We acknowledge fruitful discussions with Simon Brennecke and Manfred Lein.

\end{document}